\documentclass[english,superscriptaddress,prl,twocolumn]{revtex4}
\usepackage[toc,page,title,titletoc,header]{appendix}
\usepackage[T1]{fontenc}
 \usepackage{color}
 \usepackage{tikz}
\usepackage{enumerate}
\usepackage{booktabs}
\usepackage{bm}
\usepackage{graphicx}
\usepackage[latin1]{inputenc}
\usepackage{epstopdf}
\usepackage{amsthm}
\usepackage{amsmath}
\usepackage{shapepar}
\usepackage{color}

\usepackage{amssymb, amsbsy, amsthm}
\makeatletter

\usepackage{babel}
\makeatother
\begin{document}
\title{Universal Method to Estimate  Quantum   Coherence}

%\author{Liang-Liang Sun$^{1}$, Kishor Bharti$^{2}$, Ya-Li Mao$^{3}$,  Xiang Zhou$^{1}$, Kwek Leong Chuan$^{2,4,5}$,  Jingyun Fan$^{3}$,  Sixia Yu $^{1}${\footnote[1]{email: yusixia@ustc.edu.cn}}
%}

\author{Liang-Liang Sun}
\affiliation{Hefei National Research Center for Physical Sciences at the Microscale, University of Science and Technology of China, Hefei,
Anhui 230026, China}

\author{Sixia Yu \footnote[1]{email: yusixia@ustc.edu.cn}}
\affiliation{Hefei National Research Center for Physical Sciences at the Microscale, University of Science and Technology of China, Hefei,
Anhui 230026, China}

\date{\today{}}
\begin{abstract}
Coherence is a defining property of quantum theory that accounts for quantum advantage in many  quantum information tasks. Although many coherence quantifiers have been introduced
in various contexts, the lack of efficient methods to estimate them restricts their applications. In this
paper, we tackle this problem by proposing one universal method to provide measurable bounds for
most current coherence quantifiers. Our method is motivated by the observation that the distance
between the state of interest and its diagonal parts in reference basis, which lies at the heart of
the coherence quantifications, can be readily estimated by disturbance effect and uncertainty of the
reference measurement. Thus, our method of bounding coherence provides a feasible and broadly applicable avenue for detecting coherence, facilitating its further practical applications.
 \end{abstract}


\maketitle
\section{I. Introduction}
Coherence, captured by the  superposition principle,  is a defining property of quantum theory.   It underscores almost  all  the  quantum features,  such as   symmetry~\cite{0Extending},  entanglement~\cite{PhysRevLett.115.020403, PhysRevLett.117.020402, PhysRevA.97.022342}, quantum correlation~\cite{PhysRevA.98.032317, PhysRevLett.116.160407}.   Coherence  also   accounts for quantum advantages in various quantum information processing tasks, such as    quantum metrology~\cite{2011Advances, Giorda_2017, PhysRevLett.122.040503} and quantum cryptography~\cite{PhysRevA.51.1863, PhysRevA.99.062325}. Within a strictly  mathematical framework of resource theory,   the significance of coherence as a resource  has been fully appreciated in recent years.  Many aspects of it,  ranging from characterization~\cite{RevModPhys.89.041003},   distillation, and, catalytic~\cite{PhysRevA.94.052336, PhysRevLett.117.030401,  RevModPhys.91.025001, PhysRevLett.113.150402},  have been investigated, along with  intense analysis of how coherence plays a role in fundamental physics (see ~\cite{2018Geometric} for a review).

Quantifying coherence lies in the heart of coherence resource theory~\cite{PhysRevLett.97.220404, PhysRevLett.113.170401, RevModPhys.89.041003, PhysRevLett.116.120404, PhysRevA.93.032136,  PhysRevA.95.042337, coherencefidelity}.  Till now,  many methods have been proposed. The most initiative method is based on state distance, for example,  quantifying coherence with  the minimal distance between the state of interest and the closest coherence free state.     The typical  examples are  the relative
entropy of coherence and the $l_{1}-$norm of coherence~\cite{PhysRevLett.116.120404}. Coherence may also  be  quantified with the distance between the concerned state and its  diagonal parts in the reference basis~\cite{PhysRevLett.118.060502}. One example is the coherence of trace-norm~\cite{PhysRevA.102.022420, 2017}. Another method is via the convex roof measure. That is,   provided a quantifier for the pure state, a general mixed state's  quantifier is constructed  via a roof construction, which method leads to  the formation of coherence~\cite{PhysRevA.92.022124} and infidelity coherence measure\cite{coherencefidelity}. There are also other quantifiers such as the  robustness of coherence~\cite{PhysRevLett.116.150502},  and the Wigner-Yanase skew information of coherence~\cite{PhysRevA.95.042337}.

While many theoretical works have been devoted to a systematical research of  coherence~\cite{PhysRevLett.113.140401, PhysRevA.94.060302,   2018Geometric,  RevModPhys.89.041003},    it remains a difficult problem to efficiently estimate coherence in experiments, which  limits the applications of the quantifications  as  common tools for quantum information processing.    Clearly, one can    perform state tomography and then calculates  quantifiers  with  the derived quantum density matrix,  or estimate coherence by  employing normal witness technique~\cite{PhysRevA.93.042107, PhysRevLett.118.020403, PhysRevLett.120.230504}  and with numerical optimizations~\cite{PhysRevLett.120.170501}.  These methods   suffer from the complexities of  mathematics and  experiment setup thus lacking efficiency.
 Another  method is based on spectrum estimation~\cite{PhysRevA.99.062310},  which commonly  needs a few test measurements  to  obtain a non-trivial estimation.  These  methods, unfortunately,  are commonly restricted for estimating the  convex roof quantifiers, such as  the coherence of formation, convex roof  of  infidelity,  and convex roof of the Wigner-Yanase  skew information.

To improve the evaluation of coherence in experiments, we report  one simple and feasible detection  method, which provides both the upper and  the lower bounds  in terms of the reference measurement's uncertainty and disturbance effect, respectively. We find that coherence quantifiers are closely related to the distance between the state of interest and its diagonal parts( written in the reference basis). The distance can be upper-bounded in terms of uncertainty according to recent uncertainty-disturbance relations (UDR)~\cite{ucb,sun2022disturbance} and lower-bounded according to data processing inequality. These bounds are formulated with statistics from a universal experiment scheme.  By the method, almost all the current  coherence quantifiers of great interests are immediately  bounded as long as  one universal experiment setup output statistics. The  quantifiers include the  relative entropy of  coherence measure,  the coherence of formation,  the $l_{1}-$norm of coherence, the $l_{2}-$ norm of coherence, the  trace-norm of  coherence,  the convex roof  of infidelity coherence, the Wigner-Yanase  skew information of coherence and the convex roof construction,  and the robustness of coherence.   Thus the method exhibits merits of simplicity and broad applicability.

The rest of the paper is structured as follows. In section I, we briefly review state-distance based coherence quantifiers, then  provide  framework for upper-bounding and lower-bounding them. In section II, we provide  measurable bounds for most current coherence quantifiers,  including the distance-based quantifiers   and some other quantifiers of general interests. In section III, we compare  our method with the one based on spectrum estimation,  showing that our method is  experimentally less demanding and more efficient.

\section{II. The Framework of Detecting Coherence}
%which far extend the its initial meaning as capability of exhibiting interference visibility. According to coherence resource theory,    This framework enable one do study coherence  of the diagonal terms of  its significance  has been far extended and    been quantified in various contexts.   with  measures such as relative entropy~\cite{PhysRevLett.113.140401},  the $l_1$ norm of coherence~\cite{2018, 2017},  the robustness of coherence~\cite{PhysRevLett.116.150502}.    and quantitative account of  its connection with other fundamental concepts such as steering, nonlocality, are frequently investigated.
\subsection{A.The distance-based quantifications of coherence}
Coherence  is a quantity characterized  with respect to one  prefixed reference basis denoted by $\{|j\rangle\}$ with the revelent measurement being referred to as reference measurement.  The free states  are the ones of the form    $\sigma=\sum_{j} p_{j}|j\rangle\langle j|$. Otherwise, a non-diagonal   state contains coherence. Coherence is commonly quantified with state distance, for example, with   the minimal distance between the state of interest  and the closest incoherent  one~\cite{RevModPhys.91.025001}
\begin{eqnarray}
\mathcal{C}(\rho):=\min_{\sigma\in \mathcal{I}}\mathcal{D}(\rho, \sigma), \label{dis}
	\end{eqnarray}
where    $\mathcal{I}$ denotes  the  set of   coherence free states,  and $\mathcal{D}(\rho, \sigma)$ specifies   state distance. Apply Eq.\eqref{dis} to  the state  distances of  relative entropy, $l_{1}-$norm,the Tsallis relative $\alpha-$ entropies,  one can define quantifiers  meeting all the  criteria of the coherence resource theory~\cite{RevModPhys.89.041003}. They are referred to as coherence measures. If the chosen distance measures are  trace-norm and  fidelity~\cite{PhysRevA.91.042120, PhysRevA.93.012110},  Eq.\eqref{dis} defines coherence monotones that meet   a specific  subset of the criteria of the coherence resource theory.

The computability  of quantifier  $\mathcal{C}(\rho)$ is generally hard  except  the state distance, for which,  $\mathcal{C}(\rho)= \mathcal{D}(\rho, \rho_{d})$,  where $\rho_{d}$ specifies  the diagonal parts of $\rho$  in the reference basis.    One may simply define an   easily computable quantifier  as  the distance between $\rho$ and $\rho_{d}$~\cite{PhysRevLett.118.060502}:
 \begin{eqnarray}
\tilde{\mathcal{C}}(\rho):=\mathcal{D}(\rho, \rho_{d}),    \label{diag}
\end{eqnarray}
which  can be understood  as the  disturbance   caused  by reference measurement in $\rho$ as $\rho_{d}$  is just  the post-measurement state.    For some distance measures, Eq.\eqref{diag} defines a better monotone than the one defined  by Eq.\eqref{dis}.  One example is  trace-norm~\cite{PhysRevA.102.022420}, for which,  Eq.\eqref{diag} defines a     quantifier that   can  satisfy  more criteria (than the defined by Eq.(\ref{dis}))  and also allows a physically  well-motivated  interpretation as the capability to exhibit interference visibility~\cite{2017}.
 $\tilde{\mathcal{C}}(\rho)$ does not involve a minimization process. Therefore,
 \begin{eqnarray}
 \tilde{\mathcal{C}}(\rho)\geq \mathcal{C}(\rho). \nonumber
\end{eqnarray}

 One may also define coherence quantifier  via a convex roof technique~\cite{PhysRevA.92.022124}. That is, provided a quantifier for  pure state, one can  define   a mixed state's quantifier  via a  convex roof construction. For example,  one may define the pure state coherence  via Eq.\eqref{dis},  then the  convex roof construction  is
   \begin{eqnarray}
\mathcal{C}'(\rho):=\textstyle\min_{\{f_{i}, |\phi_{i}\rangle\}}\sum_{i} f_{i}\cdot \mathcal{C}(\phi_{i}), \label{con}
\end{eqnarray}
where the minimization is taken  over  all  the possible pure state decompositions of $\rho=\sum f_{i}|\phi_{i}\rangle\langle \phi_{i}|$.
This definition has its advantage, e.g., when
applied to  infidelity,   Eq.\eqref{con} can define  a  measure~\cite{coherencefidelity} for coherence  while Eq.\eqref{dis}  defines only a  monotone.

The above definitions  have led to many quantifiers and  also  induce the bounds for the quantifiers defined in other ways.  In the following, we first introduce a  framework for bounding them.
\begin{figure}
\centering
\includegraphics[scale=1]{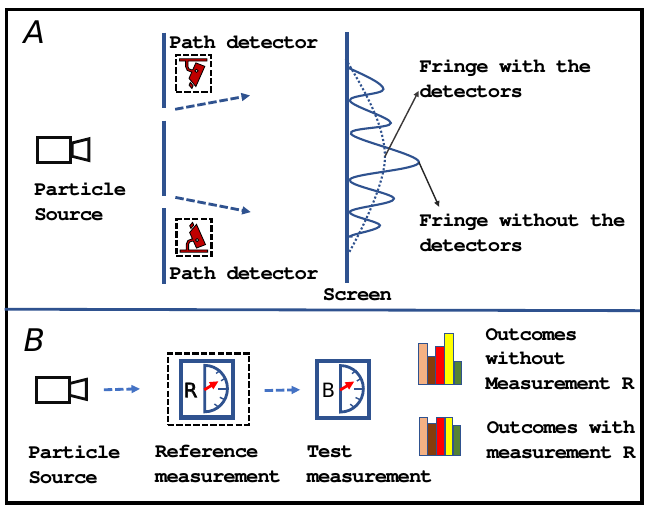}
\caption{Two-slit experiment and coherence detection protocol. Particles with and without coherence leads to different statistics when subjected to a test measurement. } \label{fig:temp}  %of
\end{figure}
\subsection{B. The framework of detecting coherence}
Detecting the nonclassical properties, such as entanglement, coherence, and randomness, is to lower-bound them  using the statistics   coming from experiments.
For  coherence, it is instructive to recall its early illustration based on the  two-slit experiment  which is shown in Fig.1A, where
the  reference measurement  is the path detectors  that erases the coherence between  paths, and the screen is an incompatible  measurement that  verifies the coherence   in terms of  the change of  interference fringes due to the destroy of coherence.  Here, we would like to  exploit how  this  idea applies to many other coherence quantifiers.  We consider   a similar  measurement setting as shown in Fig1.B. It consists of one reference measurement $\mathcal{R}$ and one following test measurement $\mathcal{B}$.  The reference measurement updates  an input state $\rho$ into an incoherent state, $i.e.$, its diagonal parts $\rho_{d}$. It then is subject to  a following measurement $\mathcal{B}=\{|b_{j}\rangle\langle b_{j}|\}$,  giving rise to a distribution $\mathbf{q'}=\{q'_{j}=\operatorname{tr}(\rho_{d}\cdot|b_{j}\rangle\langle b_{j}|)\}$. This is a typical sequential measurement scheme that can be readily   realised with off-the-shelf instruments~\cite{PhysRevLett.94.220405,PhysRevLett.112.020402, PhysRevLett.110.220402, PhysRevLett.113.250403}.   If without the measurement  $\mathcal{R}$, directly performing $\mathcal{B}$ on $\rho$ yields a distribution  $\mathbf{q}=\{q_{j}=\operatorname{tr}(\rho\cdot |b_{j}\rangle\langle b_{j}|)\}$.  Distance  between $\mathbf{q}$ and $\mathbf{q}'$ can  be understood as the disturbance introduced by the reference measurement in $\mathcal{B}$.  One may choose $\mathcal{B}$ as the one maximally incompatible with the reference measurement, $i.e.$,  $\forall i, j$,  $|\langle i|b_{j}\rangle|^{2}=\frac{1}{d}$  with $d$ specifying  dimension of the  revelent Hilbert's space.  This setting commonly can ensure a significant distance between $\mathbf{q}$ and $\mathbf{q}'$, which is in favour of coherence estimation.
We also note that  $\mathbf{q}'$ actually does not require  performing a real test measurement   after the reference measurement.   As  $\rho_{d}$ is determined by the reference measurement's distribution  as $\rho_{d}=\sum_{i}p_{i}|i\rangle\langle i|$ and $\mathbf{p}=\{p_{i}=\operatorname{tr}(\rho\cdot|i\rangle\langle i|)\}$, one can directly calculate $\mathbf{q}'$ via Born's rule. For example, the probability when $\mathcal{B}:=\{|b_j\rangle\langle b_{j}|\}$ is $q'_{i}=\operatorname{tr}(\rho_{d}\cdot|b_{i}\rangle\langle b_{j}|)=\sum_{i}c_{ij} p_{i}$   is given by  $\mathbf{p}$, where  $c_{ij}=|\langle i|b_{j}\rangle|^{2}$. In this way, only two independent measurements, namely, $\mathcal{R}$ and $\mathcal{B}$, are sufficient for giving the  statistics $\mathbf{p}$, $\mathbf{q}$, and $\mathbf{q}'$. In the following, the estimation of coherence quantifiers only involves these distributions.

 %where coherence is characterized with difference between  interference pattern corresponding to  particles with and without coherence. %In optics, this idea can be  demonstrated equivalently with the Mach-Zehnder Interferometer, and coherence is quantified with interfence visibility. These two quantifications actually capture state distance between the between concerned state and its diagonal parts
%In the two slits experiment, coherence is erased  by  a path detector settled in the slits and the screen serves as the measurement $\mathcal{B}$.   The difference reveal the difference between  distributions of a measurement specified as $\mathcal{B}$ performed on concerned state $\rho$ and the one after erising coherence, namely, $\rho_{d}$.

  %replace the path detector with the reference measurement and the screen with a arbitrary measurement $\mathcal{B}$.  $\rho_{d}$ is the post measurement state after reference  measurement, which can be prepared with sequential measurements schemes in the studies of error-disturbance relations.  Performed on $\rho$ and $\rho_{d}$ independently, two distributions specified by $\mathbf{q}$ and $\mathbf{q}'$ are obtained.  One does not necessary perform a realistic measurement.   Then one actually needs to perform one reference measurement and a test measurement $\mathcal{B}$.   We first provide a lower-bound which account for various  measures.

\emph{Lower-bounding coherence}
 First, the coherence quantifiers  of the form Eq.(\ref{diag}) can be estimated according to  data processing inequality, which states that distance between states, say $\rho$ and $\rho_{d}$,  are no less than the corresponding  classical distance between statistics coming from another measurement, say,  $\mathcal{B}$,  performed on them:
  %With a general measurement, $\mathcal{B}$,  performed  independently  on   $\rho$ and $\rho_{d}$, one obtains distributions denoted by $\mathbf{q}$, $ \mathbf{q}'$, respectively.      By  data processing inequality,   the statistics distance between  $\mathbf{q}$ and $\mathbf{q}'$ denoted with $\operatorname{D}[\mathbf{q}, \mathbf{q}']$,  lower-bounds state distance  $\mathcal{D}(\rho, \rho_{d})$
 \begin{equation}
\begin{aligned}
 \mathcal{D}(\rho, \rho_{d})\geq\operatorname{D}[\mathbf{q}, \mathbf{q}']. \nonumber
\end{aligned}
\end{equation}
 Immediately, we obtain a lower bound for the coherence measure
   \begin{equation}
\begin{aligned}
 \tilde{\mathcal{C}}(\rho)\geq \operatorname{D}[\mathbf{q}, \mathbf{q}']. \label{lower}
\end{aligned}
\end{equation}
 We highlight a useful property of the classical distance, namely,  the  convexity of  classical distance $\operatorname{D}$,
 $$ \textstyle\sum_{i}f_{i}\cdot\operatorname{D}[\mathbf{q}_{i}, \mathbf{q}'_{i}]\geq\operatorname{D}[\sum_{i}f_{i}\cdot\mathbf{q}_{i}, \sum_{i}f_{i}\cdot\mathbf{q}_{i}'],$$
which will be used  for bounding  $\mathcal{C}'(\rho)$.

Second, we consider the  convex roof-based coherence quantifier. We note that for pure state a coherence quantifier $\mathcal{C}'(\phi)$ is always a function of  distribution $\mathbf{p}_{\phi}$. This is because the diagonal elements of a pure state  are sufficient to  determine coherence quantifier as they determine a pure state up to some  relative phases.  These phases are  inessential in quantifying  coherence as they    can be modified freely with reversible incoherent operation of phase shifting. Noting that the maximum coherent state $|\phi\rangle=\frac{1}{d}\sum_{i}|i\rangle$  and   the zero-coherence pure state  $|i\rangle\langle i|$ exhibit the maximum and the minimal, respectively, uncertainty of reference measurement in the give basis. It is therefore reasonable to assume that the coherence quantifier $\mathcal{C}'(\phi)$ for pure state is  positively related  with  uncertainty $\delta_{\mathcal{D}}(\mathbf{p})$ (the subscript means that uncertainty can be related to state distance $\mathcal{D}$) and  may  be lower-bounded  with  uncertainty  or a function of it (whose definition is left in the next section).      We find that, if a pure state coherence $\mathcal{C}'(\phi)$ allows a lower bound in terms of a convex and monotonically  increasing function  of $\delta_{\mathcal{D}}(\mathbf{p}_{\phi})$ specified by  $g(\delta_{\mathcal{D}})$,  the constructed convex roof-based coherence quantifier $\mathcal{C}'(\rho)$ can be lower-bounded  as
  \begin{equation}
\begin{aligned}
 \mathcal{C}'(\rho)\geq g( \operatorname {D}(\mathbf{q}, \mathbf{q}')). \label{crc}
\end{aligned}
\end{equation}
It needs to be stressed that  $\mathcal{C}'(\rho)\geq g( \delta_{\mathcal{D}}(\mathbf{p}))$  generally does not hold for a mixed state $\rho$ due to the concavity of  $\delta_{\mathcal{D}}$, namely,   $\mathcal{C}'(\rho)\geq \sum_{i}f_{i}\cdot g(  \delta_{\mathcal{D}}(\mathbf{p}_{i}))\geq g( \sum_{i} f_{i} \cdot \delta_{\mathcal{D}}(\mathbf{p}_{i}))$ while $g( \sum_{i} f_{i} \cdot \delta_{\mathcal{D}}(\mathbf{p}_{i}))\leq  g( \delta_{\mathcal{D}}(\mathbf{p}))$, where $\sum_{i}\mathbf{p}_{i}=\mathbf{p}$ and $\sum_{i}f_{i}\cdot \mathbf{p}_{i}=\mathbf{p}$. The key idea behind Eq.(\ref{crc}) is to relax a concave uncertainty measure $\delta_{\mathcal{D}}$  into a convex disturbance measure $\operatorname {D}(\mathbf{q}, \mathbf{q}')$ using   UDRs~\cite{sun2022disturbance} stating that   one measurement's uncertainty in terms of, say $\delta_{\mathcal{D}}(\mathbf{p})$, is no less than  its disturbance effect in the measured state $\rho$  and in the subsequent test measurement $\mathcal{B}$:
  $$\delta_{\mathcal{D}}(\mathbf{p})\geq \mathcal{D} (\rho, \rho_{d})\geq \operatorname{D} (\mathbf{q}, \mathbf{q}'). $$
Then, we have $g(\sum_{i}f_{i}\cdot \operatorname {D}(\mathbf{q}_{i}, \mathbf{q}'_{i}))\geq g(\operatorname {D}(\mathbf{q}, \mathbf{q}'))$ with $\sum_{i}f_{i}\cdot \mathbf{q}_{i}=\mathbf{q}$ and $\sum_{i}f_{i}\cdot \mathbf{q}'_{i}=\mathbf{q}'$, leading to a lower-bound for coherence quantifiers.  We left the proof  of Eq.(\ref{crc}) in SM. It can be seen that the function of $g(\cdot)$ provides a way of finding the lower-bound of coherence measure in terms of disturbance $\operatorname{D} (\mathbf{q}, \mathbf{q}')$.   In the next section, we shall show that  $g(\cdot)$ can always be found  for the existing  convex roof-based  coherence quantifiers.

\emph{Estimation of upper-bounds}
In general, upper bounds for quantum properties are not as useful as the lower bounds since they may be way larger than the actual value and thus are commonly ignored in theory and experiment. Here, we can obtain the estimation of the upper bound with the outcome distribution of the reference measurement for free, i.e., without introducing extra experiment settings, and most importantly the resulting upper bound may assist the estimation of coherence in our framework.

It follows from the UDRs that the upper bounds of  the quantifiers  of the form   Eq.\eqref{dis} or Eq.\eqref{diag} are given as
\begin{equation}
\delta(\mathbf{p})\geq \tilde{\mathcal{C}}(\rho)\geq \mathcal{C}(\rho). \label{disup}
\end{equation}

The  upper bound of $\mathcal{C}'(\rho)$ is given as
\begin{equation}
\delta (\mathbf{p})\geq \sum_{i}f_{i}\cdot\delta(\mathbf{p}_{i})\geq\sum_{i}f_{i}\cdot\mathcal{C}(\phi_{i})
=\mathcal{C}'(\rho), \label{uppercon}
\end{equation}
Thus,  one reference  measurement  is sufficient for  upper-bounding the three kinds of  coherence  quantifiers.

 %With $\bf{P}(\rho)$ we specify distribution from  reference measurement performed on $\rho$  and $\bf{Q}(\rho)$ and $\bf{Q}'(\rho_{d})$ the distribution from measurement $\mathcal{B}$ performed on $\rho$ and $\rho_{d}$.  By $\{f_{i}, |\phi_{i}\rangle\}$ we denote  the decomposition of $\rho$.  Immediately, $\sum_{i}f_{i}\cdot\bf{P}(\phi_{i})=\bf{P}(\rho)$ and $\sum_{i}f_{i}\cdot\bf{Q}(\phi_{i})=\bf{Q}_{\rho}$ and $\sum_{i}f_{i}\cdot\bf{Q}'(\phi_{i,d})=\bf{Q}'(\rho_{d})$. Specifically, we will abbreviate $\bf{P}(\rho)$, $\bf{Q}(\rho)$,  and $\bf{Q}'(\rho_{d})$ as $\bf{P}$, $\bf{Q}$, and $\bf{Q}'$.   Generally, we impose no assumption on $\mathcal{B}$ unless noted.
Till now, with the distributions of $\mathbf{p}$,   $\mathbf{q}$,  and $\mathbf{q}'$, both the upper and the lower bounds are obtained. A possible large gap between them roughly indicates that: (i) the state of interest contains little coherence, and (ii) the setting of $\mathcal{B}$ is not well-chosen. Then the lower bound may be optimized by choosing other settings of $\mathcal{B}$ or one may almost confirm that the state of interest contains little coherence. In this way the upper bound assists the estimation of lower bound.

\section{III.  Detecting Coherence in Various contexts}
In the following, we use the above  framework to estimate  coherence quantifiers having general interests.
\subsection{A. The relative entropy of  coherence measure and the coherence of formation }
First, we consider the  relative entropy of coherence~\cite{PhysRevLett.113.140401} and the coherence of formation~\cite{aberg2006quantifying, PhysRevLett.97.220404, PhysRevA.92.022124}, which are defined by applying Eq.\eqref{dis} and Eq.\eqref{con} to  the   relative entropy
  $$S(\rho\|\sigma):=\operatorname{Tr}(\rho\log\rho- \rho\log \sigma).$$
Relative entropy of coherence is  a legitimate measure. It has operational meaning as  asymptotic coherence distillation rate~\cite{PhysRevLett.116.120404} and also quantifies   the quantum randomness  under the quantum adversaries (with
independent measurements)~\cite{PhysRevA.92.022124, 2018rand, 2019rand}. The quantifier is defined as
  $$\mathcal{C}_{re}(\rho):=\min_{\sigma \in \mathcal{I}}S(\rho\|\sigma)=S(\rho\|\rho_{d}).$$
The coherence of formation has an  interpretation of   the asymptotic coherence dilution
rate~\cite{PhysRevLett.116.120404}. It   also quantifies the quantum randomness  under the classical adversaries (with
independent measurements)~\cite{PhysRevA.92.022124, 2018rand, 2019rand}. The quantifier  reads
   $$\mathcal{C}'_{re}(\rho):=\min_{\{f_{i},\phi_{i}\}}\sum_{i} f_{i} \cdot S(\phi_{i}\|\phi_{i, d}).$$
 The  UDR corresponding to the relative entropy   is given as
  $$\operatorname{H}(\mathbf{p})\geq S(\rho\|\rho_{d})\geq \operatorname{H}(\mathbf{q}\|\mathbf{q}'), $$
where  the Shannon entropy  $\operatorname{H}(\mathbf{p}):=-\sum_{i}p_{i}\log p_{i}$ defines measurement uncertainty  $\delta_{re}(\mathbf{p})$ and the relative entropy $S(\rho\|\rho_{d})$ defines  disturbance in quantum state and the classical relative entropy    $\operatorname{H}(\mathbf{q}\|\mathbf{q}'):=\sum_{i}q_{i}\log q_{i}- \sum_{i}q_{i}\log q'_{i}$ defines the disturbance in measurement $\mathcal{B}$ denoted with $\rm D_{re}(\mathbf{q}, \mathbf{q}')$.
Based on the general arguments just provided, we immediately have~\cite{sun2022disturbance}
\begin{equation}
\operatorname{H}(\mathbf{p})\geq \mathcal{C}_{re}(\rho), \quad \mathcal{C}'_{re}(\rho)\geq \operatorname{H}(\mathbf{q}\|\mathbf{q}'). \label{defc}
\end{equation}
For $\mathcal{C}_{re}(\rho)$, the bounds are obvious. For $\mathcal{C}'_{re}(\rho)$, we have  $\mathcal{C}'_{re}(|\phi\rangle)=\operatorname{H}(\mathbf{p}_{\phi})$, which leads to a definition of  convex and monotonically increasing $g$ function  as $g(x)=x$. Then a lower bound for $\mathcal{C}'_{re}(\rho)$, namely,   $\operatorname{H}(\mathbf{q}\|\mathbf{q}')$,  follows from  Eq.(\ref{crc}) and uncertainty disturbance relation~\cite{sun2022disturbance}.

\subsection{B.  The $l_{1}-$norm,  the $l_{2}-$ norm, and the trace-norm of  coherence}
\emph{The $ l_{1}-$norm of coherence.} The  $l_{1}-$norm of coherence  quantifies  the maximum entanglement that can be created from coherence  under incoherent operations acting on the system and an
incoherent ancilla~\cite{PhysRevA.97.022342}. It  has been used  to investigate  the speed-up of quantum computation~\cite{PhysRevA.93.012111, PhysRevA.95.032307},  wave-particle duality~\cite{2018Geometric, PhysRevA.92.012118, PhysRevA.92.042101}, and uncertainty principle~\cite{PhysRevA.96.032313}. The quantifier is defined via Eq.(\ref{dis}) as~\cite{PhysRevLett.113.140401}
$$\mathcal{C}_{l_{1}}(\rho)=\min_{\sigma \in \mathcal{I}}\mathcal{D}_{l_{1}}(\rho, \sigma)=\mathcal{D}_{l_{1}}(\rho, \rho_{d}).$$
where the $l_{1}-$norm
$$\mathcal{D}_{l_{1}}(\rho, \sigma')=\textstyle\sum_{i, j}|\rho_{ij}-\sigma'_{ij}|$$
with   $\rho_{ij}$ and $\sigma'_{ij}$ specifying    matrixes' elements.

The  UDR corresponding to the $l_{1}-$norm distance is
 $$ \|\mathbf{p}\|_{\frac{1}{2}}-1\geq \mathcal{D}_{l_{1}}(\rho, \rho_{d}),$$
 where $\|\mathbf{p}\|_{x}=(\sum_{i} p^{x}_{i})^{\frac{1}{x}}$.   The inequality  is because that $\sum_{i\neq j}|\rho_{ij}|\leq \sum_{i\neq j}\sqrt{p_{i}p_{j}}=\|\mathbf{p}\|_{\frac{1}{2}}-1 $.
Based on Eq.(\ref{disup}) and the UDR, the  $\mathcal{C}_{l_{1}}(\rho)$ is estimated via
\begin{equation}
   \|\mathbf{p}\|_{\frac{1}{2}}-1 \geq \mathcal{C}_{l_1}(\rho)\geq 2|\mathbf{q}-\mathbf{q}'|, \label{cl1}
\end{equation}
where $|\mathbf{q}-\mathbf{q}'|:=\frac{1}{2}\sum_{i}|q_{i}-q'_{i}|$ is the Kolmogorov distance.  The lower bound side is due to
$\sum_{i\neq j}|\rho_{ij}|= \sum_{i> j}$$\operatorname{tr}|\varrho_{ij}|\geq \operatorname{tr}|\rho-\rho_{d}|
\geq 2 |\mathbf{q}-\mathbf{q}'|$, where $\operatorname{tr}|A|=\operatorname{tr}\sqrt{AA^{\dagger}}$ and  $\varrho_{ij}:=|j\rangle\langle j|\rho|i\rangle\langle i|+|i\rangle\langle i|\rho|j\rangle\langle j|$ and   $\sum_{i\neq j}\operatorname{tr}|\varrho_{ij}|\geq \operatorname{tr}|\rho-\rho_{d}|$ is due to the convexity.

\emph{The $l_{2}-$norm of coherence.}
The $l_{2}-$norm of coherence~\cite{PhysRevLett.117.030401, RevModPhys.89.041003} has an operational  interpretation  as state uncertainty~\cite{PhysRevA.103.042423} and is   also  employed  to study non-classical correlations~\cite{Sun_2017}. The quantifier reads
$$\mathcal{C}_{l_{2}}(\rho):=\min_{\sigma \in \mathcal{I}} \mathcal{D}_{l_{2}}(\rho, \sigma)=\mathcal{D}_{l_{2}}(\rho, \rho_{d}),$$
where the $l_{2}-$norm or   the (squared) Hilbert-Schmidt distance
$\mathcal{D}_{l_{2}}(\rho, \sigma):=\operatorname{tr}(\rho-\sigma)^{2}.$

The  UDR corresponding to  $l_{2}-$norm is
$$1-\|\mathbf{p}\|^{2}_{2}\geq\mathcal{D}_{l_{2}}(\rho, \rho_{d}),$$
which  is due to $\operatorname{tr}(\rho-\rho_{d})^{2}$$= \operatorname{tr}(\rho^{2}-\rho\cdot\rho_{d})\leq 1-\|\mathbf{p}\|^{2}_{2}$.
Based on Eq.\eqref{disup} and Eq.\eqref{lower},  $\mathcal{C}_{l_{2}}(\rho)$ is bounded as
\begin{eqnarray}
1-\|\mathbf{p}\|^{2}_{2} \geq \mathcal{C}_{l_2}(\rho)\geq \|\mathbf{q}-\mathbf{q}'\|^{2}_{2}, \label{cl2}
\end{eqnarray}
where  the lower bound is due to data processing inequality with  measurement $\mathcal{B}$ being required to be  projective~\cite{OZAWA2000158}.

\emph{Trace-norm of coherence.} The trace-norm of coherence has an interpretation of interference visibility and reads~\cite{PhysRevA.102.022420}
\begin{eqnarray}
\tilde{\mathcal{C}}_{tr}(\rho)=\rm \mathcal{D}_{tr}(\rho, \rho_{d}) \nonumber
\end{eqnarray}
where  the distance of  trace-norm $\mathcal{D}_{tr}(\rho, \sigma)=\textstyle\rm\frac{1}{2} tr|\rho-\sigma|.$
According to   UDR corresponding to  this distance~\cite{sun2022disturbance}
 \begin{eqnarray}
\sqrt{1-\|\mathbf{p}\|^{2}_{2}}\geq \rm \mathcal{D}_{tr}(\rho, \rho_{d}),  \nonumber
\end{eqnarray}
we have
\begin{eqnarray}
\sqrt{1-\|\mathbf{p}\|^{2}_{2}}\geq \tilde{\mathcal{C}}_{tr}(\rho)\geq |\mathbf{q}-\mathbf{q}'|.
\end{eqnarray}
Again, the lower bound is due to data processing inequality.

\subsection{C. Convex roof coherence of  infidelity}
Now, we consider the convex roof coherence of infidelity~\cite{coherencefidelity}
 $$\mathcal{C}'_{if}(\rho):=\textstyle \min_{\{f_{i}, |\phi_{i}\rangle \}}\sum_{i} f_{i}\cdot \mathcal{C}_{if}(\phi_{i}),$$
where the pure state coherence is quantified as
  $$\mathcal{C}_{if}(\phi)=\min_{\sigma\in \mathcal{I}}\mathcal{D}_{if}(\phi,\sigma)$$
  with  the infidelity
 $\mathcal{D}_{if}(\rho,\sigma):=\sqrt{1-\rm F(\rho, \sigma)}$ and
  $\operatorname{F}(\rho, \sigma)=[\rm Tr(\sqrt{\sqrt{\rho}\sigma\sqrt{\rho}})]^{2}$.
 The UDR corresponding to infidelity is~\cite{sun2022disturbance}.
  $$\sqrt{1- \|\mathbf{p}\|_{2}^{2} }\geq \mathcal{D}_{if}(\rho, \rho_{d}). $$
By   Eq.(\ref{uppercon}), the coherence measure acquires  an upper bound as $\sqrt{1- \|\mathbf{p}\|_{2}^{2} }$.

To  lower-bound the measure using Eq.\eqref{crc}, we note that $\mathcal{C}_{if}(\phi)\geq \frac{\sqrt{2}}{2} \sqrt{1-\|\mathbf{p}\|^{2}_{2}}:= \frac{\sqrt{2}}{2} \delta_{if}(\mathbf{p})$
(See SM), which leads to a definition of $g$ function as $g(x)=\frac{\sqrt{2}}{2}x$.
By Eq\eqref{crc}, we have
 $\textstyle \mathcal{C}'_{if}(\rho)\geq\frac{\sqrt{2}}{2} \operatorname{D}_{if}(\mathbf{q}, \mathbf{q}')$, where $\operatorname {D}_{if}(\mathbf{q}, \mathbf{q}'):=\textstyle\sqrt{1-(\sum_{i} \sqrt{q_{i}q'_{i}})^{2}}.$
Thus, we finally have
  \begin{eqnarray}
\sqrt{1- \|\mathbf{p}\|_{2}^{2} } \geq \mathcal{C}'_{if}(\rho) \geq \textstyle \frac{\sqrt{2}}{2} \operatorname{D}_{if}(\mathbf{q}, \mathbf{q}')\label{ifc}.
\end{eqnarray}

\begin{table*}[tb]
	\centering
	\caption{Measurable lower bounds for coherence quantifiers.} \begin{tabular}{{l}{c}{c}{c}{c}{c}{c}{c}{c}}
  \hline\hline
Quantifier &  $\mathcal{C}_{re}$ &$\mathcal{C}_{l_{1}}$ &  $\mathcal{C}_{l_{2}}$ & $\tilde {\mathcal{C}}_{tr}$ & $\mathcal{C}'_{if}$   & $C_{s}$   & $C'_{s}$ &  $C_{rob}$ \\
 \hline
%Real value& $\mathrm{H}(\sin^{2}\frac{\theta}{2})$ &  $\sin \theta$  & $\sin \theta$ & $\sin\frac{\theta}{2}$ & $\frac{1}{2}\sin^{2}\frac{\theta}{2}$&  $\textstyle\frac{1}{2}\sin^{2}\frac{\theta}{2}$ \\
 % \hline
 Upper Bound   &$\operatorname{H}(\mathbf{p})\quad$ &   $\|\mathbf{p}\|_{\frac{1}{2}}-1 \quad$ &  $ 1-\|\mathbf{p}\|^{2}_{2}\quad$  & $\sqrt{1-\|\mathbf{p}\|^{2}_{2}}\quad$ & $\sqrt{1-\|\mathbf{p}\|^{2}_{2}}\quad$  & $1-\|\mathbf{p}\|^{2}_{2}\quad$ &$1-\|\mathbf{p}\|^{2}_{2}$ & $\|\mathbf{p}\|_{\frac{1}{2}}-1$ \\
  \hline
 Lower Bound   &$\operatorname{H}(\mathbf{q}\|\mathbf{q}')\quad$ &   $2|\mathbf{q}-\mathbf{q}'| \quad$ &  $ \|\mathbf{q}-\mathbf{q}'\|^{2}_{2}\quad$  & $|\mathbf{q}-\mathbf{q}'|\quad$ & $\frac{\sqrt{2}}{2}\operatorname{D}_{if}(\mathbf{q}, \mathbf{q}')\quad$  & $\frac{1}{2}\|\mathbf{q}-\mathbf{q}'\|^{2}_{2}\quad$ &$\operatorname{D}^{2}_{if}(\mathbf{q}, \mathbf{q}')\quad$ & $\frac{\|\mathbf{q}-\mathbf{q}'\|^{2}_{2}}{\|\mathbf{p}\|_{\infty}}$ \\
  \hline
\end{tabular}
	\label{tab1}
\end{table*}

%By the way, our method implies a trade-off between the coherence quantifier and the infidelity $\mathcal{D}_{if}(\rho, \rho_{d})$. That is,    optimizing  the  classical disturbance over $\mathcal{B}$,  $\max_{\mathcal{B}}\operatorname{D}_{if}(\mathbf{q}, \mathbf{q}')=\mathcal{D}_{if}(\rho, \rho_{d})$ leads to
%    \begin{eqnarray}
%\mathcal{C}'_{if}(\rho) \geq \textstyle \frac{\sqrt{2}}{2} \mathcal{D}_{if}(\rho, \rho_{d})\label{14}.
%\end{eqnarray}
Till now, we have estimated  many distance-based coherence  quantifiers.
In the following, we shall use the  method  to estimate  the   quantifiers  going  beyond the above definitions.  The following quantifiers  shall be specified by $C$ instead of $\mathcal{C}$ for the sake of specification.
\subsection{D.  The  coherence of  Wigner-Yanase skew information  and the convex roof construction}
Coherence can  also be  quantified based on quantum Fisher information, which is a basic concept in the field of quantum metrology that places the fundamental limit on the information accessible by performing measurement on quantum state.  Two remarkable quantifiers are the Wigner-Yanase skew  information of  coherence and the convex roof construction  based on it.

\emph{The Wigner-Yanase skew  information of  coherence.} The  Skew information coherence is a legitimate coherence measure and defined as~\cite{PhysRevA.95.042337}$$C_{s}(\rho)=\textstyle \sum^{d}_{1=j}\operatorname{I}(\rho, |j\rangle\langle j|),$$
where $\operatorname{I}(\rho, |j\rangle\langle j|)\equiv -\frac{1}{2}\operatorname{Tr}([\rho, |j\rangle\langle j|])^{2}$ represents  the Wigner-Yanase skew information subject to the projector $|j\rangle\langle j|$. This measure  has ever been estimated with  a spectrum estimation  method employing the  standard overlap measurement technique~\cite{PhysRevA.95.042337}, where one needs to perform $2d-2$ measurements to estimate the coherence of  a $d-$dimensional  system. In the following, our method can reduce the number to three.

First, we can reexpress the measure as $\textstyle C_{s}(\rho)=1-\sum_{j}\langle j|\sqrt{\rho}|j \rangle^{2} $~\cite{PhysRevA.95.042337}, then an upper bound readily  follows as
\begin{eqnarray}
\textstyle C_{s}(\rho)&=&1-\sum_{j}\langle j |\sqrt{\rho}| j \rangle^{2}\leq 1-\textstyle\sum_{j}\langle j |\rho| j\rangle^{2}\nonumber\\ &=&1-\|\mathbf{p}\|_{2}^{2}. \nonumber
\end{eqnarray}
In order to derive  a lower-bound, we use  an inequality $C_{s}(\rho)\geq \frac{1}{2}\mathcal{C}_{l_{2}}(\rho)$~\cite{PhysRevA.95.042337} whose bound  has already been bounded  in Eq.\eqref{cl2}.  Then, we  have
\begin{eqnarray}
1-\|\mathbf{p}\|_{2}^{2}\geq C_{s}(\rho)\geq \frac{1}{2}\|\mathbf{q}-\mathbf{q}'\|^{2}_{2}.
\end{eqnarray}

\emph{The Convex Roof Construction.}
 With   $C_{s}(\phi)$ quantifying coherence of a pure state $|\phi\rangle$,   the Wigner-Yanase skew  information can  lead to a convex roof construction of coherence as~\cite{PhysRevA.103.012401}
\begin{eqnarray}
C'_{s}(\rho)=\textstyle\min_{\{f_{i},\phi_{i}\}}\sum_{i}f_{i}\cdot C_{s}(\phi_{i}). \nonumber
\end{eqnarray}
This quantifier can   be equivalently  defined via  the  quantum Fisher information (up to an inessential factor) with respective to reference measurement~\cite{Yu2013,PhysRevA.103.012401}:
\begin{eqnarray}
C'_{s}(\rho)=\frac{1}{4}\sum_{j} \operatorname{F}(\rho, |j\rangle\langle j|),  \nonumber
\end{eqnarray}
where $\operatorname{F}(\rho, |j\rangle\langle j|):=\sum_{k, l}2\frac{(\lambda_k-\lambda_{l})^{2}}{\lambda_k+\lambda_{l}}|c_{kl}|^{2}$ specifies quantum information of $\rho$ subject to the measurement projector $|j\rangle\langle j|$  and $c_{kl}=\langle\phi_{k}|j\rangle\langle j|\phi_{l}\rangle$ with  $|\phi_{n}\rangle$ being  the $n$th eigenvector of $\rho$ and $\lambda_{n}$ being  the weight.

Given that $C_{s}(\phi)=1-\|\mathbf{p}(\phi)\|^{2}_{2}=\delta^{2}_{if}(\mathbf{p})$, we can define  $g(\delta_{if})=\delta_{if}^{2}$.
By  Eq.\eqref{crc}, the lower bound of $C'_{s}(\rho)$ follows  as
$$ C_{s}(\rho) \geq \operatorname{D}^{2}_{if}(\mathbf{q}, \mathbf{q'}).$$
 Due to the convexity of $1-\|\mathbf{a}\|^{2}_{2}$ with respect to distribution $\mathbf{a}$,  an upper bound is immediately  obtained   as
\begin{eqnarray}
1-\|\mathbf{p}\|^{2}_{2}\geq C'_{s}(\rho). \nonumber
\end{eqnarray}
Thus, the quantifier is bounded as
\begin{eqnarray}
1-\|\mathbf{p}\|^{2}_{2}\geq C'_{s}(\rho)\geq \operatorname{D}^{2}_{if}(\mathbf{q}, \mathbf{q'})
\end{eqnarray}

%Again, with the same argument of Eq.(\ref{14}) a computable lower-bound  of $C'_{s}(\rho)$ is given  as $\mathcal{D}^{2}_{if}(\rho, \rho_{d})$. \emph{ Specifically, the lower bound now has an independent interest as measurable bounds for quantum fisher information. This is because that, Then we have
%\begin{eqnarray}
%\sum_{j} \operatorname{F}(\rho, |j\rangle\langle j|)\geq 4\operatorname{D}^{2}_{if}(\rho, \rho')\geq 4\operatorname{D}^{2}_{if}(\mathbf{q}, \mathbf{q'}).  \nonumber
%\end{eqnarray}
%This  bridges quantum fisher information and infidelity and  sheds light on an estimation of quantum fisher information.}

\subsection{F.  The  robustness of coherence}
One important coherence monotone is the robustness of coherence~\cite{PhysRevLett.116.150502}, which quantifies the advantage enabled by a quantum state in a phase discrimination task.  For a given state $\rho$, it is defined as  the minimal mixing required to make the state incoherent
\begin{eqnarray}\nonumber
C_{Ro}(\rho)=\min_{\tau}\left\{s \geq 0 \bigg|\frac{\rho+s\tau}{1+s}:=\sigma \in \mathcal{I} \right\}
\end{eqnarray}
where $\tau$ is a general quantum state.
With the inequality~\cite{PhysRevLett.116.150502}  $$C_{Ro}(\rho)\geq \frac{\mathcal{D}_{l_{2}}(\rho,\rho_{d})}{\|\rho_{d}\|_{\infty}}. $$
where note that  $\|\rho_{d}\|_{\infty}=\|\mathbf{p}\|_{\infty}$ is just  the maximum element in  $\mathbf{p}$. It follows Eq.\eqref{cl2} that
\begin{eqnarray}
C_{Ro}(\rho)\geq \frac{\|\mathbf{q}-\mathbf{q}'\|^{2}_{2}}{\|\mathbf{p}\|_{\infty}}. \nonumber
\end{eqnarray}
With the inequality $C_{Ro}(\rho)\leq \mathcal{C}_{l_{1}}(\rho)$ and Eq.\eqref{cl1},  we have
\begin{eqnarray}
 \|\mathbf{p}\|_{\frac{1}{2}}-1 \geq C_{Ro}(\rho)\geq \frac{\|\mathbf{q}-\mathbf{q}'\|^{2}_{2}}{\|\mathbf{p}\|_{\infty}}. \label{cro}
\end{eqnarray}
The robustness of coherence has ever been detected based on the witness method~\cite{PhysRevA.93.042107, PhysRevLett.118.020403, PhysRevLett.120.230504}, where the mathematics of constructing witness and experiment setup  are generally complex.

Finally we summarize the obtained measurable lower bounds in Table.I. These bounds are all formulated in terms of the statistics $\mathbf{p}$, $\mathbf{q}$, and $\mathbf{q}'$.  As the essential quantity, the disturbance $\operatorname{D}$ can readily be measured by  performing a  sequential measurements scheme, which has been well-developed  in the study of error-disturbance relations~\cite{PhysRevLett.94.220405,PhysRevLett.112.020402, PhysRevLett.110.220402, PhysRevLett.113.250403}.   In our approach, as the lower-bounds  are  smooth functions of experimental statistics, the statistics errors due to  imperfection of implementation of measurement or state preparation  are  one order smaller comparing with the lower-bounds. These aspects  make our protocol quite feasible.

It is also of practical interests to consider  the tightness of these bounds.  We find that the bounds  of coherence measures $\mathcal{C}_{l_{2}}$ and $\tilde{\mathcal{C}}_{tr}$  can be saturated for any  input state when  the test measurement $\mathcal{B}$ is taken as the eigenvectors of $\rho-\rho_{d}$, and the bounds for the convex-roof based measure $C'_{s}$ can be saturated if the concerned state is pure, and the bound of $\mathcal{C}_{l_{1}}$ is tight in qubit case (when $\mathcal{B}$ is taken as the eigenvectors of $\rho-\rho_{d}$).   The bounds for other quantifiers either cannot not be non-trivially  saturated or only be saturated for some specific states.

%\subsection{F. Upper-bounding   Tsallis  $\alpha$ entropy.}
%Our method have leaded bounds for  many   coherence quantifiers of great interest. However,  show  left one open question open, namely,  estimating  the
  %coherence quantifier based on Tsallis relative $\alpha$ entropies,   which is defined as  $\operatorname{\mathcal{D}}_{\alpha}(\rho\|\sigma):=\frac{1}{1-\alpha}[1-\operatorname{Tr}
%(\rho^{\alpha}\sigma^{1-\alpha})]$ with $0\leq \alpha< 1$. The associated    quantifier is~\cite{PhysRevA.93.032136, 2018Coherence}
%\begin{eqnarray}
%C_{T}(\rho)=\min_{ \sigma \in \mathcal{I}} \operatorname{\mathcal{D}}_{\alpha}(\rho\|\sigma).\nonumber
%	\end{eqnarray}

%Another legitimate coherence measure defined via (\eqref{dis}) is defined by
%Our framework can provide an upper bound
%\begin{eqnarray}
%C_{T}(\rho)\leq \textstyle \frac{1}{1-\alpha}(1- %\|\mathbf{P}\|^{2-\alpha}_{2-\alpha}).
%	\end{eqnarray}
%It is due to
%\begin{eqnarray}\nonumber
%C_{T}(\rho)&\leq& \textstyle \operatorname{\mathcal{D}}_{\alpha}(\rho\|\rho_{d})\\ \nonumber
%&\leq& \textstyle \frac{1}{1-\alpha}[1-\operatorname{Tr}
%(\rho\rho^{1-\alpha}_{d})]\\ \nonumber
%&=& \textstyle \frac{1}{1-\alpha}(1- \|\mathbf{P}\|^{2-\alpha}_{2-\alpha}).
%	\end{eqnarray}
%Lower-bounding this coherence measure is an open question.

\section{IV. Efficiency  Argument}
The previous   coherence estimation protocols commonly  apply to only  a few coherence quantifiers. The collective measurement protocol~\cite{collective} applies to the  relative entropy of coherence and the $l_{2}-$norm of coherence. The witness method~\cite{PhysRevA.93.042107, PhysRevLett.118.020403, PhysRevLett.120.230504, PhysRevA.103.012409} applies to the robustness of coherence, the $l_{1}-$norm, and the $l_{2}-$norm of coherence.  One quite simple and efficient  method is based on spectrum estimation via majorization theory~\cite{PhysRevA.99.062310}, which can be used to estimate the $l_{1}-$norm, $l_{2}-$norm of coherence, the robustness of coherence, and the relative entropy of coherence.  This method employs a similar measurement scheme to ours. In the following, we compare our  approach with it.

\subsection{A. Comparison with the method based on spectrum estimation}
The  spectrum estimation method  is based on the theory of majorization~\cite{PhysRevA.99.062310}. A probability $\mathbf{a}$  is said to majorize  a probability distribution $\mathbf{b}$, specified as $\mathbf{a}\succeq \mathbf{b}$,  if their elements  satisfy $\sum^{k}_{i=1}a^{\downarrow}_{i}\geq\sum^{k}_{i=1}b^{\downarrow}_{i}$ $\forall k$, where the superscript means that the elements are  arranged in a descending ordered, namely, $\mathbf{a}^{\downarrow}=(a^{\downarrow}_{1}, a^{\downarrow}_{2}, \cdots a^{\downarrow}_{d})$, $\mathbf{b}^{\downarrow}=(b^{\downarrow}_{1}, b^{\downarrow}_{2}, \cdots b^{\downarrow}_{d})$ with $a_{i}\geq a_{i+1}$ and  $b_{i}\geq b_{i+1}$. Clearly, the spectrum of $\rho$, specified  as $\boldsymbol{\lambda}=(\lambda_{1}, \lambda_{2}\cdots \lambda_{d})$,  majorizes   distribution from  any projection  measurement, say $\mathcal{B}$,  performed  on $\rho$,   $i.e.$, $\boldsymbol{\lambda}\succeq\mathbf{q}$.  By the Shur convexity theorem, $\operatorname{H}(\mathbf{a})\geq \operatorname{H}(\mathbf{b})$ if  $\mathbf{b}\succeq\mathbf{a}$. Thus,  $\mathcal{C}_{r}(\rho)= \operatorname{H}(\mathbf{p})- \operatorname{H}(\boldsymbol{\lambda}) \geq \operatorname{H}(\mathbf{p})-\operatorname{H} (\mathbf{q})$ with $\mathbf{p}$ and $\mathbf{q}$ being distributions from the reference measurement and $\mathcal{B}$, which provides a nontrivial lower bound if  $ \mathbf{q} \succ\mathbf{p}$.  Generally, as state of interest  is unknown  one needs to try a few settings of $\mathcal{B}$ to ensure $ \mathbf{q} \succ\mathbf{p}$. In this paper, we deal with the disturbance effect, namely, $\operatorname{D}(\mathbf{q}, \mathbf{q}')$, which is zero iff $\rho-\rho_{d}$ is perpendicular with all  the elements of  $\mathcal{B}$ simultaneously.  The settings  resulting in such a failure lie in a space of measure zero. Therefore, our method almost always works even the $\mathcal{B}$ is chosen arbitrarily.   As a simple illustration,   assume that  $\rho$ is given  as $|\phi\rangle=\sin\frac{\pi}{8}|0\rangle+\cos\frac{\pi}{8}|1\rangle$ and the reference basis is $\{|0\rangle, |1\rangle\}$, immediately, $\mathbf{p}=\{\sin^{2}\frac{\pi}{8}, \cos^{2}\frac{\pi}{8}\}$.   By the spectrum estimation  method, a nontrivial estimation, namely,   $\operatorname{H} (\mathbf{q})-\operatorname{H}(\mathbf{p})> 0$ requires that $|\langle\phi|\mathcal{B}|\phi\rangle|> \cos^{2}\frac{\pi}{8}-\sin^{2}\frac{\pi}{8}=\frac{\sqrt{2}}{2}$. With our method,   $\operatorname{D}(\mathbf{q}, \mathbf{q}')\neq0$ requires that $\mathcal{B}\neq\sigma_{z}, \sigma_{y}$,  which is much weaker than the above requirement.
\subsection{B Numerical results for the qubit case}
The lower bounds provided in Table.I work very well.
As the second illustration of efficiency, we  consider  the qubit case.  For the sake of  computability, we let $\rho$ be a  pure state.  How well a quantifier is estimated can be naturally quantified  with the ratio of the estimate to the exact value. For each quantifier, we calculate the  average of the ratio over the randomly chosen measurement $\mathcal{B}$ and randomly chosen pure state (See SM for details) with two different methods, $i.e.$, our method and the one based on spectrum estimation.  The averages are listed in the following table, where the one based on   our method is specified by $Q^{C}_{D}$ and  by  $Q^{C}_{S}$ the other method.
\begin{table}[htbp]
	\centering
	\caption{Efficiencies  in estimating   coherence of qubit states.}
	\begin{tabular}{{l}{c}{c}{c}{c}{c}{c}{c}{c}}
  \hline\hline
 Measure &  $\mathcal{C}_{re}$ &$\mathcal{C}_{l_{1}}$ &  $\mathcal{C}_{l_{2}}$ & $\tilde {\mathcal{C}}_{tr}$ & $\mathcal{C}'_{if}$   &  $C_{s}$ &  $C'_{s}$ & $C_{rob}$ \\
 \hline
%Real value& $\mathrm{H}(\sin^{2}\frac{\theta}{2})$ &  $\sin \theta$  & $\sin \theta$ & $\sin\frac{\theta}{2}$ & $\frac{1}{2}\sin^{2}\frac{\theta}{2}$&  $\textstyle\frac{1}{2}\sin^{2}\frac{\theta}{2}$ \\
 % \hline
$Q^{C}_{D}$   &$0.36$ &   $\frac{2}{\pi}$ &  $\frac{1}{2}$ & $\frac{2}{\pi}$ & $0.22$ &$0.25$ &$0.31$ & $0.28$ \\
  \hline
$Q^{C}_{S}$  & $0.17$ & $0.29$ &  $0.20$ &$/$ & $/$ &$/$ & $/$& $0.20$ \\
  \hline
\end{tabular}
\end{table}
It can be seen that our method enjoys   wider applicability and higher efficiency.

\section{IV.  Conclusion}
In conclusion, we have provided a universal and straightforward  method to estimate  coherence. It enables us to give   measurable bounds for many  quantifiers of general interest, where all the bounds are expressed  as  functions of experiment accessible data  $\mathbf{p}$,  $\mathbf{q}$, and $\mathbf{q}'$  without involving  cumbersome mathematics. This is advantageous over the previous methods, which only applies to one or a few measures and cannot apply to  the  quantifiers  based on convex roof construction.  Our  approach exhibits many  desired features: experiment friendly,  broad applicability,  and mathematical simplicity therefore   serves as an efficient coherence detecting method.

For the   possible further  researches in the quantum foundation, we note that   disturbance effect is one basic concept in the quantum foundation that closely relates to nonlocality, uncertainty principle,  and the security of quantum cryptography. Thus,  the framework may inspire novel connection among  these concepts, for example,  nonlocality and coherence.

\emph{Acknowledgement ---} Supports from Guangdong Provincial Key Laboratory Grant No.2019B121203002 and fundings SIQSE202104 are acknowledged.

\bibliography{coherence}

\clearpage
\appendix
\newpage

\newpage
\appendix
\section{Supplemental Material.}
\subsection{A. Proof of Eq.(\ref{crc})}
For any  pure state ensemble, for example,   the one achieving   the minimal of convex roof-based measure, specified by $\rho=\sum_{i}f_{i}\cdot |\phi_{i}\rangle\langle\phi_{i}|$, we have
$\sum_{i}f_{i}\cdot\mathbf{q_{i}}= \mathbf{q} $ and $\sum_{i}f_{i}\cdot\mathbf{q_{i}}'= \mathbf{q}' $.
For $\mathcal{C}'(\rho)$, we have
\begin{eqnarray}
\mathcal{C}'(\rho)&= &\sum_{i} f_{i}\cdot\mathcal{C}(\phi_{i})
\geq \sum_{i} f_{i}\cdot g(\delta_{\mathcal{D}}(\mathbf{p}_i))\nonumber \\
 &\geq&
 \sum_{i} f_{i}\cdot g(\operatorname{D}(\mathbf{q}_{i}, \mathbf{q}'_{i}))
 \geq g( \sum_{i} f_{i}\cdot \operatorname{D}(\mathbf{q}_{i}, \mathbf{q}'_{i})) \nonumber\\
 &\geq &g(  \operatorname{D}(\mathbf{q}, \mathbf{q}')). \label{proof1}
\end{eqnarray}
The first inequality is due to the definition of $g$ function. The second inequality is due to  UDRs, which  relax a concave quantity  $\delta_{\mathcal{D}}(\mathbf{p})$ into a convex one  $ \operatorname{D}(\mathbf{q}, \mathbf{q}')$ . The third is due to  the convexity of $g(\cdot)$. The fourth is due to  the  convexity  of the classical distance  $\operatorname{D}$ and the   monotonicity assumption of $g(\cdot)$.

\section{B. Proof of Eq.(\ref{ifc})}
To lower-bound the coherence measure, let us first consider a pure state case, for example, $|\phi\rangle=\sum_{i}\sqrt{p_{i}(\phi)}e^{i\psi_{i}}|i\rangle$, we have
 \begin{eqnarray}
\mathcal{C}_{if}(\phi)=\min_{\sigma \in \mathcal{I}}\mathcal{D}_{if}(\rho,\sigma) =\sqrt{1- p} \nonumber
\end{eqnarray}
 where $p=\max_{i}\{|p_{0}(\phi)|, \cdots, |p_{d-1}(\phi)|\}$.
  Note that,  $$ \sqrt{2(1-p)}\geq\textstyle \sqrt{(1+p)(1- p) }\geq \sqrt{1- \|\mathbf{p}(\phi)\|_{2}^{2} }.$$
The UDR corresponding to infidelity  is
  $$\sqrt{1- \|\mathbf{p}\|_{2}^{2} }\geq \operatorname {D}_{if}(\mathbf{q}(\phi), \mathbf{q}'(\phi_{d})). $$
   where $\operatorname {D}_{if}(\mathbf{q}(\phi), \mathbf{q}'(\phi_{d})):=\sum_{j} \sqrt{q_{j}q'_{j}}$ is  classical infidelity.
Then we have  $\mathcal{C}_{if}(\phi)\geq \frac{\sqrt{2}}{2} \sqrt{1-\|\mathbf{p}\|^{2}_{2}}:= \frac{\sqrt{2}}{2} \delta_{if}(\mathbf{p})$, which leads to a definition of $g$ function as $g(x)=\frac{\sqrt{2}}{2}x$. The lower bound of $\mathcal{C}'_{if}(\rho)$ follows from Eq(\ref{proof1})
  \begin{eqnarray}
\mathcal{C}'_{if}(\rho)\geq \textstyle \frac{\sqrt{2}}{2} \rm D_{if}(\mathbf{q}, \mathbf{q}').  \nonumber
\end{eqnarray}

To obtain an upper bound, we use  Eq.(7) and the UDR $\rm D_{if}(\rho, \rho_{d})\leq \sqrt{1- \|\mathbf{p}\|_{2}^{2}} $ then  have
  \begin{eqnarray}
\mathcal{C}'_{if}(\rho)\leq \textstyle \sum_{i} f_{i}\cdot \sqrt{1- \|\mathbf{p}_{i}\|^{2} }\leq  \sqrt{1- \|\mathbf{p}\|_{2}^{2}}\nonumber
\end{eqnarray}
where we have used the concavity of $\sqrt{1- \|\mathbf{p}\|_{2}^{2}}$.

\section{C. Tightness argument}
\begin{table*}[!htb]
\centering
\begin{tabular}{{l}{c}{c}{c}{c}{c}{c}{c}{c}{c}}
  \hline\hline
 Measure &  $\mathcal{C}_{re}$ &$\mathcal{C}_{l_{1}}$ &  $\mathcal{C}_{l_{2}}$ & $\mathcal{C}'_{if}$  &$\tilde {\mathcal{C}}_{tr}$ &  $C'_{s}$ & $C_{s}$& $C_{rob}$ \\
 \hline
 Exact value $C(\phi)$& $\quad \mathrm{H}(\sin^{2}\frac{\theta}{2})$ &  $\quad \sin \theta$  & $\quad\frac{1}{2}\sin^{2} \theta$ & $\quad\min\{\sin\frac{\theta}{2}, \cos\frac{\theta}{2}\}$ & $\quad\frac{1}{2}\sin^{2}\theta$& $\quad\frac{1}{2}\sin^{2}\theta$& $\quad\textstyle\sin\theta$  &\quad $\sin\theta$\\
\hline
$Q^{C}_{D}$   &$0.36$ &   $\frac{2}{\pi}$ &  $\frac{1}{2}$& $\frac{2}{\pi}$   & $0.22$ &$0.25$ & $0.31$ & $0.28$ \\
  \hline
$Q^{C}_{M}$  & $0.17$ & $0.29$ &  $0.20$ &$/$ & $/$ &$/$ &$/$ &  $0.20$ \\
$Q^{'C}_{M}$  & $0.20$ & $0.32$ &  $0.22$ &$/$ & $/$ &$/$ &$/$ &  $0.22$ \\
 % \hline
  \hline
\end{tabular}
\caption{Comparing the efficiencies of our method and the one based on the spectrum estimation. }
\end{table*}
For the sake of computability,  we consider an arbitrary pure state $\rho=|\phi\rangle\langle\phi| $ with $|\phi\rangle=\sin\frac{\theta}{2}|0\rangle +\cos \frac{\theta}{2}e^{i\psi}|1\rangle$ and  $0\leq\theta\leq \pi$ and $0\leq\psi\leq 2\pi$.   The exact value of the  coherence   is specified by $C(\phi)$.  We choose the  test  measurement $\mathcal{B}$  as the one maximally incompatible with  the reference measurement. Its setting  thus is determined  up to a relative phase as   $\{\frac{1}{\sqrt{2}}|0\rangle+\frac{e^{i\psi'}}{\sqrt{2}}|1\rangle; \frac{1}{\sqrt{2}}|0\rangle-\frac{e^{i\psi'}}{\sqrt{2}}|1\rangle\}$, and none of them is of  priority.  We average the ratio $\frac{L^{C}_{D,\mathcal{B}}}{C(\phi)}$  over all the possible   $\mathcal{B}$ as
$$\frac{\bar{L}^{C}_{D,\mathcal{B}}}{C(\phi)}:=\frac{1}{2\pi}\int^{2\pi}_{0} \frac{L^{C}_{D,\mathcal{B}}}{C(\phi)} d\psi',$$   where one  lower bound    $L^{C}_{D,\mathcal{B}}$ is given under the choice of    $\mathcal{B}$.
As state of interest  is unknown, we  access  the average  performance of the protocol
 over  all the  pure states as $\frac{\bar{L}^{C}_{D,\mathcal{B}}}{C(\phi)}$.
$$Q^{C}_{D}:=\frac{1}{4\pi}\int^{2\pi}_{0}\int^{\pi}_{0} \frac{\bar{L}^{C}_{D,\mathcal{B}}}{C(\phi)} \sin\theta d\theta d\psi.$$
The $Q^{C}_{D}$ for the various quantifiers are calculated and listed in the following table. Using the above maximally incompatible  measurement implies a greater disturbance for $\mathcal{C}_{l_{1}}$,   $\mathcal{C}_{l_{2}}$,   $\tilde {\mathcal{C}}_{tr}$,     $C_{s}$, and $C_{rob}$,  than using a  random choice of the test measurement $\mathcal{B}$ along direction $\vec{B}=\{\sin\alpha \sin\psi'', \sin\alpha \cos \psi'', \cos \alpha\}$ with $0\leq\alpha\leq \pi$ and $0\leq\psi''\leq 2\pi$. This is because typical quantities such as $|\mathbf{q}-\mathbf{q}'|$ or  $\|\mathbf{q}-\mathbf{q}'\|^{2}_{2}$  involved in lower bounds, which are $|\sin \alpha \sin \theta \cos (\psi'-\psi'') |$ or $\frac{1}{2}|\sin \alpha \sin \theta \cos (\psi'-\psi'') |^{2}$, respectively,  attain their maximums at $\alpha=\frac \pi2$ and $\psi'=\psi''$. We also calculate  $Q^{C}_{D}$ for the other three quantifiers, namely,  $C_{re}$, $\mathcal{C}'_{if}$, and $C'_{s}$ using general choice of $\mathcal{B}$, and obtain the average lower bounds as  $0.266$, $0.365$, and $0.234$.

We also calculate the average performance $Q^{C}_{M}$ using the method based on the spectrum estimation. One key difference is that the test measurement $\mathcal{B}$ is not need to be the one maximum incompatible with the reference measurement. This is because in the protocol based on spectrum estimation the closer the setting of  $\mathcal{B}$ to the eigenvectors of state of interest, the better the estimate is. However,   the state of interest is unknown and so do its eigenvectors. Therefore there  is no reasonable to choose the test measurement $\mathcal{B}$ as the one maximum incompatible with the reference measurement.      According to  this protocol, $\mathcal{B}$ is taken randomly  as  $\mathcal{B}=\vec{\sigma}\cdot\vec{B}$ and $\vec{B}=\{\sin\alpha \sin\psi'', \sin\alpha \cos \psi'', \cos \alpha\}$ with $0\leq\alpha\leq \pi$ and $0\leq\psi''\leq 2\pi$ and $\vec{\sigma}$ being  the Pauli matrix. The  average performance is defined as
$$Q^{C}_{M}:=\frac{1}{4\pi}\int^{2\pi}_{0}\int^{\pi}_{0} \frac{\bar{L}^{C}_{M,\mathcal{B}}}{C(\phi)} \sin\theta d\theta d\psi.$$
 with $\frac{\bar{L}^{C}_{M,\mathcal{B}}}{C(\phi)}$ being defined as
 $$\frac{\bar{L}^{C}_{M,\mathcal{B}}}{C(\phi)}
 :=\frac{1}{4\pi}\int^{2\pi}_{0}\int^{\pi}_{0} \frac{L^{C}_{M,\mathcal{B}}}{C(\phi)} \sin\alpha d\alpha d\psi''.$$
We have calculated the coherence quantifiers that this method is applicable to. As a comparison, we also calculate the performance when $\mathcal{B}$ taking the measurement  maximally incompatible with the reference measurement, which is specified with $Q^{'C}_{M}$.
By the table, it is shown that our method enjoys higher efficiency and wider applicability.

\end{document}